\documentstyle{mn}
\input epsf

\def\msun{{M_\odot}}
\def\beq{\begin{equation}}
\def\eeq{\end{equation}}
\def\bey{\begin{eqnarray}}
\def\eey{\end{eqnarray}}
\def\beqarray{\begin{eqnarray}}
\def\eeqarray{\end{eqnarray}}

\def\Dd{D_{\rm d}}

\def\Ds{D_{\rm s}}
\def\Dds{D_{\rm ds}}

\def\d{ {\rm d}}

\def\apj{{ApJ}}
\def\apjl{{ApJ}}
\def\aj{{AJ}}
\def\mnras{{MNRAS}}
\def\aap{{A{\&}A}}

\def\bv{\mbox{\boldmath $v$}}
\def\bx{\mbox{\boldmath $x$}}
\def\bd{\mbox{\boldmath $d$}}

\def\ri{r_{\rm i}}
\def\re{r_{\rm e}}

\def\spose#1{\hbox to 0pt{#1\hss}}
\def\lta{\mathrel{\spose{\lower 3pt\hbox{$\sim$}}
    \raise 2.0pt\hbox{$<$}}}
\def\gta{\mathrel{\spose{\lower 3pt\hbox{$\sim$}}
    \raise 2.0pt\hbox{$>$}}}

\title[Fitting Gravitational Lenses]        
{Fitting Gravitational Lenses: Truth or Delusion}

\author[N. Wyn Evans and Hans J. Witt]
{N. Wyn Evans$^{1,2}$ and Hans J. Witt$^3$ \\
$^1$ Theoretical Physics, 1 Keble Rd, Oxford, OX1 3NP \\
$^2$ Institute of Astronomy, Madingley Rd, Cambridge, CB3 0HA \\
$^3$ Baumkamp 8, 22299 Hamburg, Germany }

\pubyear{2002}

\begin{document}

\label{firstpage}

\maketitle
\begin{abstract}
The observables in a strong gravitational lens are usually just the
image positions and sometimes the flux ratios. We develop a new and
simple algorithm which allows a set of models to be fitted exactly to
the observations. Taking our cue from the strong body of evidence that
early-type galaxies are close to isothermal, we assume that the lens
is scale-free with a flat rotation curve. External shear can be easily
included. Our algorithm allows full flexibility as regards the angular
structure of the lensing potential. Importantly, all the free
parameters enter linearly into the model and so {\it the lens and flux
ratio equations can always be solved by straightforward matrix
inversion}. The models are only restricted by the fact that the
surface mass density must be positive. 

We use this new algorithm to examine some of the claims made for
anomalous flux ratios. It has been argued that such anomalies betray
the presence of substantial amounts of substructure in the lensing
galaxy.  We demonstrate by explicit construction that some of the lens
systems for which substructure has been claimed can be well-fit by
smooth lens models. This is especially the case when the systematic
errors in the flux ratios (caused by microlensing or differential
extinction) are taken into account. However, there is certainly one
system (B\,1422+231) for which the existing smooth models are definitely
inadequate and for which substructure may be implicated.

Within a few tens of kpc of the lensing galaxy centre, dynamical
friction and tidal disruption are known to be very efficient at
dissolving any substructure. Very little substructure is projected
within the Einstein radius.  The numbers of strong lenses for which
substructure is currently being claimed may be so large that this
contradicts rather than supports cold dark matter theories.
\end{abstract}


\section{INTRODUCTION}

The fitting of models to observational data has always been a major
concern in strong gravitational lensing (e.g., Schechter 2000). In
most cases, the lens is only composed of 2 (a ``doublet'') or 4 (a
``quadruplet'') point-like images. For the doublets, the observable
constraints are the four relative coordinates of the images with
respect to the lensing galaxy, together with the one flux ratio of the
first image to the second. For the quadruplets, this becomes eight
relative coordinates and three flux ratios. Only for a handful of lens
systems are time delays available.  It therefore happens very
frequently that lens models have more degrees of freedom than the
number of constraining observations.  In fact, the situation is often
even worse than we have just described. Sometimes the centre of the
lensing galaxy itself cannot be reliably identified. Sometimes a lens
system is complicated by the possible effects of nearby bright
galaxies. Very often, the flux ratios are untrustworthy, either
because of differential extinction in the optical bands (e.g., Falco
et al. 1999) or because of microlensing in the optical and radio
(e.g., Irwin et al. 1989; Koopmans \& de Bruyn 2000) or because of
scintillation, scatter-broadening and free-free absorption in the
radio (e.g., Patnaik et al. 1992, Jones et al. 1996, Winn, Rusin \&
Kochanek 2003). It is therefore already clear that considerable
caution must be exercised in interpreting the results of fits to
gravitational lens systems.

The approach of modellers has been largely two-pronged. First, general
non-parametric methods have been developed (e.g., Saha \& Wiliams
1997). These have proved powerful in exploring the range of
degeneracies in the lensing mass that can give rise to a particular
image configuration. For example, in the case of PG\,1115+080, Saha \&
Williams present three different models obeying the lensing
constraints, but for which the lensing mass distribution resembles a
face-on spiral, an edge-on disk or a flattened elliptical galaxy
respectively. Second, many simple parametric models have been
thoroughly explored, such as those based on elliptically stratified
potentials (e.g., Witt 1996; Witt \& Mao 1997) or densities (e.g.,
Kassiola \& Kovner 1993; Mu{\~n}oz, Kochanek \& Keeton 2001). The
advantage of this is that the models may already incorporate some of
the known properties of nearby galaxies.  However, this approach may
also be dangerous because simple ansatze like elliptical potentials or
surface mass densities can introduce unexpected properties. These
properties may be so severe that, for example, the flux ratios may not
be well fitted and wrong conclusions may be drawn. This can be seen
most clearly in the case of elliptical potentials, in which there are
strong constraints on the flux ratios (e.g., Witt \& Mao 2000; Hunter
\& Evans 2001).

All this attains added significance in the light of recent claims of
evidence for substructure from ``anomalous flux ratios'' in strong
lensing. Mao \& Schneider (1998) were the first to point out that
substructure may be needed to explain the flux ratios in some cases.
The instance that they selected, the quadruplet B\,1422+231, has three
highly magnified bright images and one much fainter image.  Three
highly magnified images occur generically near a cusp, and a Taylor
expansion gives a universal relationship that the sum of the fluxes of
the two outer images should equal the flux of the middle image. This
is strongly violated in B\,1422+231 leaving substructure on top of the
smooth model as the believable culprit. This result is supported by
the detailed models of B\,1422+231 by Bradac et al. (2002) which find
that the discrepancy requires substructure on the mass scale $\sim
10^6 \msun$.  Recently, Dalal \& Kochanek (2002) looked at seven
predominantly radio quadruplets and argued that the flux ratios were
anomalous by comparison with those expected for simple isothermal
lenses with external shear.  They claimed that the anomalous flux
ratios implied the existence of $\sim 2$ per cent of the mass of the
lensing galaxy in substructure.  Metcalf \& Zhao (2002) looked at five
optical quadruplets and similarly claimed evidence for anomalous flux
ratios on comparison with those expected for simple elliptical
power-law potentials with shear.  The problem with this procedure is,
of course, that anomalous flux ratios may not be the result of
substructure at all, but may simply reflect deficiencies in the
modelling.

This motivates us to introduce in sections 2 and 3 a new approach for
fitting which can incorporate the flux ratios at outset and which can
permit an arbitrary azimuthal dependence for the surface density.
Importantly, all the free parameters enter linearly into the models
and so {\it the lens and flux ratio equations can always be solved by
straightforward matrix inversion}. The models are only restricted by
the fact that the surface mass density must be positive. This
algorithm is used in section 4 to assess the evidence for anomalous
flux ratios. Readers who are mainly interested in this application may
skip section 3 entirely on the first reading. This section is rather
mathematical and explicitly derives the linear equations for the
parameters to be fitted.
   
\begin{figure*}
 \epsfysize=7.5cm \centerline{\epsfbox{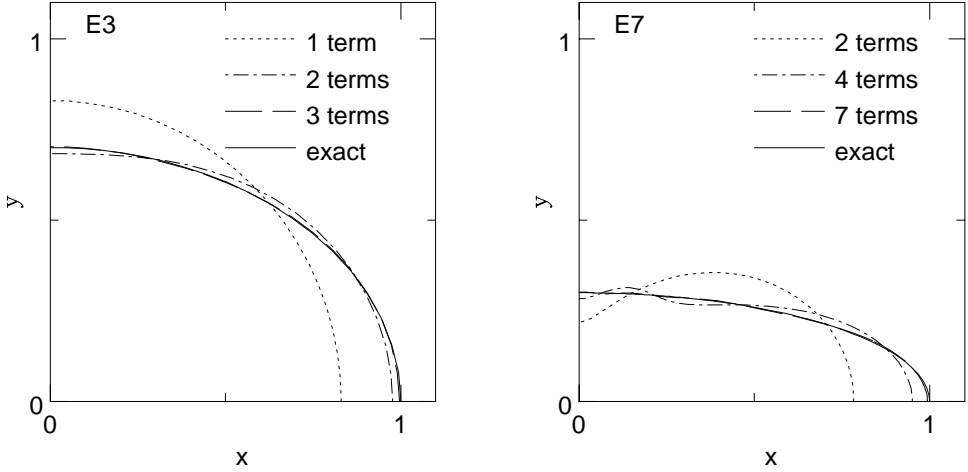}}
\caption{The two plots show truncated Fourier series approximations to
the equidensity contours of an exactly elliptical E3 (left panel) and
an E7 (right) galaxy. For moderate flattenings such as E3, excellent
results are obtained with only the first three non-vanishing terms in
the expansion. Even for the most highly flattened configurations such
as E7, the approximation becomes very accurate when the first seven
non-vanishing terms are included.  } \label{fig:sample}
\end{figure*}

\section{SCALE-FREE MODELS}

\subsection{Potential and Surface Mass Density}

Scale-free galaxy models with arbitrary power-law fall-off are widely
used in galactic astronomy and dynamics (e.g., Toomre 1982; Evans
1994; Evans, Carollo \& de Zeeuw 2000).  The isophotes of a scale-free
galaxy have the same shape at every radius and are completely
described by a {\em shape function} $G(\theta)$, which depends only on
the position angle $\theta$ with respect to the major axis.  In such
galaxies, the convergence $\kappa$ and deflection potential $\phi$ are
proportional to a power of $r$, namely
\begin{equation}
\kappa = {1 \over 2} G(\theta) r^{\beta-2}, \qquad\qquad \phi 
= r^\beta F(\theta).
\label{eq:scalefree}
\end{equation}
Here, ($r,\theta$) are familiar polar coordinates in the lens plane.
The power is restricted to lie in the range $0 < \beta < 2$ for galaxy
models. For $\beta < 0$, the mass at the centre of the galaxy does not
converge.  For $\beta >2 $, the surface mass density increases with
radius, which is unrealistic. The important $\beta =1$ case
corresponds to an everywhere flat rotation curve.

In gravitational lensing, some simple scale-free potentials were
already investigated by Kassiola \& Kovner (1993). They were followed
by Witt, Mao \& Keeton (2000), who pointed out that the time delay is
independent of the angular function $F(\theta)$ for the special case
$\beta = 1$.  Many applications to this general model followed (see
Evans \& Witt 2001; Zhao \& Pronk 2001; Wucknitz 2002; Cardone et
al. 2002). 

From Poisson's Law, we have
\begin{equation}
G(\theta) = \beta^2 F(\theta) + F^{\prime\prime} (\theta).
\label{eq:ode}
\end{equation}
Given $F(\theta)$, it is straightforward to generate $G(\theta)$.
However, it is also possible to solve this ordinary differential
equation for $F(\theta)$ in terms of $G(\theta)$, as shown in Appendix
A. This is useful, as it gives the scale-free potential corresponding
to a given set of scale-free equidensity contours.

The critical curves can be given analytically. Using the determinant
of the Jacobian, we obtain
\begin{eqnarray}
\label{eq:critcurve}
\det J (r,\theta) &=& (\beta -1 ) r^{2\beta-4} [ \beta F
F^{\prime\prime}- (\beta -1) F^{\prime 2} + \beta^2 F^2 ] \nonumber \\
& & - r^{\beta -2 } [F^{\prime\prime} +\beta^2 F] +1 = 0.
\label{eq:detjay}
\end{eqnarray}
Setting $\rho = r^{\beta -2}$ yields a quadratic equation in $\rho$
which can easily be solved. The two roots correspond to the radial and
the tangential critical curves respectively. They can be mapped onto
the radial and tangential caustics with the lens equation using the
components of the deflection angle~(\ref{eq:defangle}).  For the flat
rotation curve case ($\beta =1$), the equation for the critical curve
becomes linear and thus the caustic network is readily established, as
shown in Evans \& Witt (2001).

\subsection{Fourier Expansions}

Since $F(\theta)$ and $G(\theta )$ are periodic with period $2\pi$, we
can expand them as Fourier series (see e.g. Bronshtein \& Semendyayev
1998).  For convenience, we start with the potential $F(\theta)$ and
write
\begin{equation}
\label{eq:Fourier}
F(\theta) = { a_0 \over 2 } + \sum_{k = 1}^{\infty} [a_k \cos(k
\theta) + b_k \sin(k \theta)].
\end{equation}
Using eq.(\ref{eq:ode}), we obtain now for $G(\theta)$
\begin{equation}
\label{eq:Fourieragain}
G(\theta) = { a_0 \beta^2\over 2 }\!+\!\sum_{k = 1}^{\infty} [a_k 
(\beta^2\!-\!k^2) \cos(k\theta)\!+\!b_k (\beta^2\!-\!k^2) \sin(k
\theta)].
\label{eq:pois}
\end{equation} 
The coefficients in the Fourier series must obey the condition
$G(\theta) \geq 0 $ for all $\theta$ to obtain a positive surface mass
density.  After constructing a fit, this is easy to check {\it a
posteriori}.

As an example, let us consider an elliptical surface mass density with
a flat rotation curve ($\beta =1$), which has
\begin{equation}
G(\theta) = A (\cos^2\theta + q^{-2} \sin^2\theta)^{-1/2}.
\end{equation}
Since $G(\theta) = G(-\theta)$ is an even function, all $b_k$ must
vanish ($b_k = 0$). Since $G(\theta) = G(\pi-\theta)$, all $a_k$ with
$k$ equal to an odd integer must also vanish. So, the only terms that
remain in the Fourier series are the even integer cosine coefficients
(cf. also Appendix B).  Figure~\ref{fig:sample} shows how few terms
are really required for accurate approximation. For an E3 galaxy, a
Fourier series truncated after the first three non-vanishing terms
already gives an excellent approximation to the true equidensity
contours. Even for a highly flattened E7 galaxy, only the first seven
non-vanishing terms are needed. This demonstrates that the Fourier
expansions can give good results with few terms.

The advantage of the Fourier expansion is that it makes it very easy
to solve the lens equation. For this reason, we list the Fourier
coefficients for the popular elliptic density and potential models
with flat rotation curves in Appendix B (although we do not need to
use them in the main body of this paper).

\subsection{The Image Positions and Fluxes}

The lens equation relates the position of the source ($\xi,\eta$)
to the positions of the image ($x,y$) through the derivative of the 
lensing potential (e.g., Schneider, Ehlers \& Falco 1992)
\begin{equation}
\xi = x - {\partial \phi \over \partial x}, \qquad\qquad
\eta = y - {\partial \phi \over \partial y}.
\end{equation}
Using polar coordinates in the image plane, the lens equation can be
recast as
\begin{eqnarray} \label{eq:xi}
\xi &=& r \cos\theta - r^{\beta-1} [ \beta  \cos\theta F(\theta) - \sin\theta
F^{\prime} (\theta)], \\ \label{eq:eta}
\eta &=& r \sin\theta - r^{\beta-1} [ \beta  \sin\theta F(\theta) + \cos\theta
F^{\prime} (\theta) ],
\end{eqnarray}
where we can express $F(\theta)$ and $F^{\prime}(\theta)$ by Fourier
series. Let us assume that we have measured the image positions
$(r_\ell, \theta_\ell)$ (in polar coordinates) of a gravitational lens
and inserted them into the lens equation.  Immediately, we notice that
all unknown quantities (i.e. $\xi$, $\eta$, and all $a_i$ and $b_i$)
enter linearly into the lens equation, except for $\beta$.

Let us now add in the constraint that the magnification ratios of the
images are equal to the observed values. Suppose that the ratio of the
magnification of the $\ell$th image to the $k$th is measured to be
$f_{k\ell}$, then
\begin{equation}
f_{k\ell} = {\det J (r_\ell,\theta_\ell) \over \det J (r_k, \theta_k)},
\end{equation}
where $\det J$ is given in eq.~(\ref{eq:detjay}).  We observe that the
coefficients $a_i$ and $b_i$ enter in simple mixed quadratic form into
the equations.  However, for the special case of a flat rotation curve
($\beta = 1$), all unknown quantities enter linearly into the equation
for the magnification relation.

In other words, {\it in the astrophysically important flat rotation
curve case, the Fourier coefficients describing the lensing galaxy are
related to the observables by a simple matrix equation}. This makes
the problem of fitting to both the image positions and the flux ratios
a straightforward matter of matrix inversion.

A number of recent investigations have exploited the well-known
linearity of the lens equation to expand the potential in basis
functions (e.g., Keeton 2001).  Such methods can be a powerful way of
exploring a large set of models, although the drawback is that some of
the parameter space is often associated with the creation of
additional images. The need to check for this can undermine the
utility of starting with such an algorithm. It is worthwhile
emphasising how our work differs from such expansions.  First, the
fundamental and new contribution of our paper is that the flux ratio
equation is linear for a galaxy with a flat rotation curve.
Therefore, in our algorithm both the lens equations and the flux ratio
equations are simultaneously solved. Second, the caustic network is
straightforwardly available using the work in Section 2.1, and so it
is very easy to check for the creation of unwanted additional images.

\begin{table*}
\begin{center}
\begin{tabular}{lccccccc} \hline
\null & $-\Delta \alpha$ & $\Delta \delta$ & Radio fluxes 
& Mid-infrared flux & H band & I band \\ 
\null & (in ${}^{\prime\prime}$)
& (in ${}^{\prime\prime}$) & (in $\mu$ Jy) & fractions & (in mags)
& (in mags) \\ \hline
A & $0.0$             & $0.0$             & $65.5\pm 8.4$  
& $0.27 \pm 0.02$ & $14.96 \pm 0.06$ & $15.92 \pm 0.12$ \\ 
B & $0.673\pm 0.003$  & $1.697 \pm 0.003$ & $64.2 \pm 8.4$ 
& $0.30\pm 0.02$ & $15.46 \pm 0.02$ & $17.21 \pm 0.11$\\ 
C & $-0.635\pm 0.003$ & $1.210 \pm 0.003$ & $26.5\pm 8.4$  
& $0.16 \pm 0.02$ & $15.71 \pm 0.03$ & $16.77 \pm 0.12$\\ 
D & $0.866\pm 0.003$  & $0.528 \pm 0.003$ & $59.4 \pm 8.4$ 
& $0.27 \pm 0.02$& $16.00 \pm 0.04$ & $17.39 \pm 0.04$ \\ 
G & $0.075 \pm 0.004$ & $0.939 \pm 0.003$ & \null & $\null$ \\ \hline
\end{tabular}
\end{center}
\caption{Observational data on the Einstein Cross. The optical
positions are taken from the CASTLES survey, as are the H and I band
fluxes.  The radio fluxes are provided by Falco et al. (1996). The
mid-infrared flux fractions are given in Agol, Jones \& Blaes (2000).
}
\label{table:ecross}
\end{table*}
\begin{table*}
\begin{center}
\begin{tabular}{cccccccccccc} \hline
Band & $\xi$ & $\eta$ & $a_0$ & $a_2$ & $b_2$ & $a_3$ & $b_3$ 
& $a_4$ & $b_4$ & $a_5$ & $b_5$ \\ \hline
Radio & 0.0698 & $-0.0134$ & 1.7748& $-0.0422$ & 0.0430 & 0.0004 & $-0.0017$
& 0.0007 & 0.0012 & 0.0000 & 0.0008 \\
Mid-IR & 0.0662 & $-0.0131$ & 1.7711 & $-0.0417$ & 0.0417 & 0.0000 &
0.0002 & 0.0009 & 0.0005 & $-0.0002$ & 0.0001 \\
I & 0.0689 & $-0.0142$ & 1.7670 & $-0.0368$ & 0.0477 & $-0.0043$ & 0.0018 &
$-0.0016$ & $-0.0018$ & 0.0029 & 0.0029 \\
H & 0.0640 & $-0.0132$ & 1.7718 & $-0.0392$ & 0.0448 & $-0.0013$ & 0.0011 &
0.0002 & $0.0004$ & 0.0019 & 0.0025 \\ \hline
\end{tabular}
\end{center}
\caption{Fourier coefficients of the solutions for the Einstein
Cross. They correspond to the bold lines in Figure~\ref{fig:ecross},
for which the positions and flux ratios are reproduced within the
errors and are given in the same coordinate system as the data in
Table~\ref{table:ecross} but translated to the lens centre.}
\label{table:modelsone}
\end{table*}
\begin{figure*}
 \epsfysize=12.cm \centerline{\epsfbox{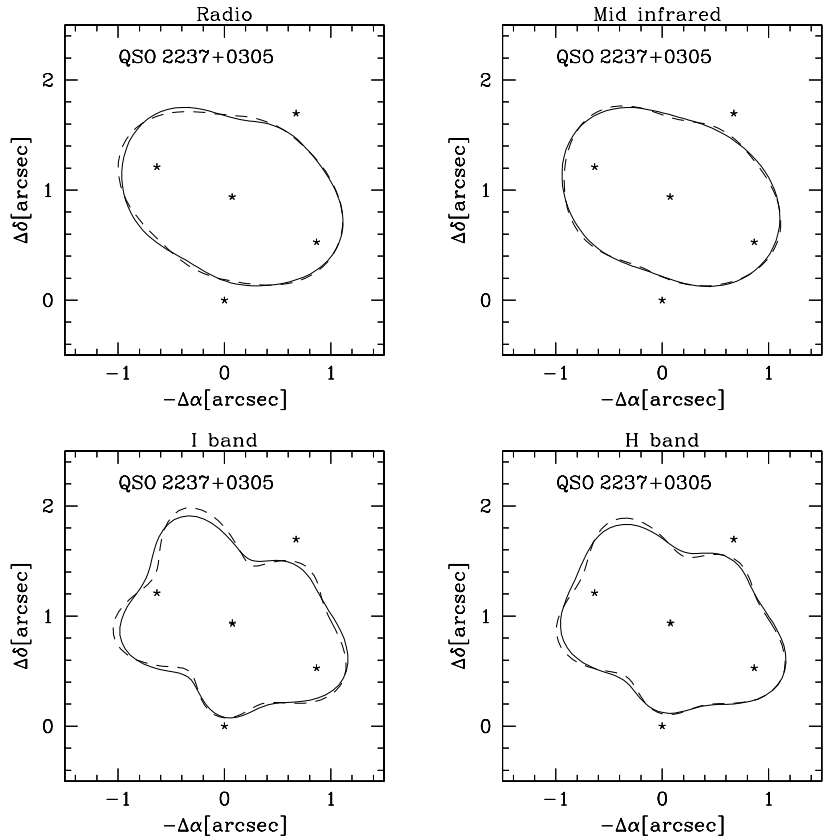}}
\caption{Each panel shows in a dashed line the critical curve (which
is also an equidensity contour when the rotation curve is flat) of a
lens model for the Einstein Cross that exactly fits the image
positions and the flux ratios. The full curve shows the equidensity
contour of a model that fits the data within the
uncertainties. Working clockwise from the top left, the panels use the
flux ratios in the radio, mid infrared, H band and I band
respectively. The locations of the four images and the lensing galaxy
are also marked.}
\label{fig:ecross}
\end{figure*}
\begin{table*}
\begin{center}
\begin{tabular}{lcc|cc|cc} \hline
\null & \multicolumn{2}{c}{$-\Delta \alpha$ (in ${}^{\prime\prime}$)}& 
 \multicolumn{2}{c}{$\Delta \delta$ (in ${}^{\prime\prime}$) } & 
  \multicolumn{2}{c}{Radio Fluxes} \\
\null & Predicted & Observed & Predicted & Observed 
& Predicted & Observed \\ \hline
A & $0.0$ & $0.0$ & $0.0$ & $0.0$ & $65.5$ & $65.5$  \\
B & $0.673$ & $0.673$ & $1.697$ & $1.697$ & $61.1$ & $64.2$ \\
C & $-0.635$ & $-0.635$ & $1.210$ & $1.210$ & $27.9$ & $26.5$  \\
D & $0.866$ & $0.866$  & $0.528$ & $0.528$ & $56.6$ & $59.4$ \\ 
G & $0.070$ & $(0.075)$  & $0.939$ & $(0.939)$ & $-$ & $-$ \\ \hline
\null & \multicolumn{2}{c}{$-\Delta \alpha$ (in ${}^{\prime\prime}$)}& 
 \multicolumn{2}{c}{$\Delta \delta$ (in ${}^{\prime\prime}$) } & 
  \multicolumn{2}{c}{Mid-infrared flux fractions} \\
\null & Predicted & Observed & Predicted & Observed 
& Predicted & Observed \\ \hline
A & $0.0$ & $0.0$ & $0.0$ & $0.0$ & $0.27$ & $0.27$  \\
B & $0.673$ & $0.673$ & $1.697$ & $1.697$ & $0.27$ & $0.30$ \\
C & $-0.635$ & $-0.635$ & $1.210$ & $1.210$ & $0.15$ & $0.16 $  \\
D & $0.866$ & $0.866$  & $0.528$ & $0.528$ & $0.30$ & $0.27$ \\ 
G & $0.075$ & $(0.075)$  & $0.939$ & $(0.939)$ & $-$ & $-$ \\ \hline
\end{tabular}
\end{center}
\caption{Predicted and observed quantities for two of the solutions of
the Einstein Cross given in Figure~\ref{fig:ecross}. The upper (or
lower ) panel corresponds to the lens model with equidensity contour
given by the bold curve in the panel for the radio (or mid-infrared
fluxes).  The positions $-\Delta \alpha$ and $\Delta \delta$ are
computed relative to image A.  Notice that all the data are
well-reproduced within even the formal uncertainties and the flux
ratios are not anomalous.}
\label{table:ecrossresults}
\end{table*}
\begin{figure*}
 \epsfysize=12.cm \centerline{\epsfbox{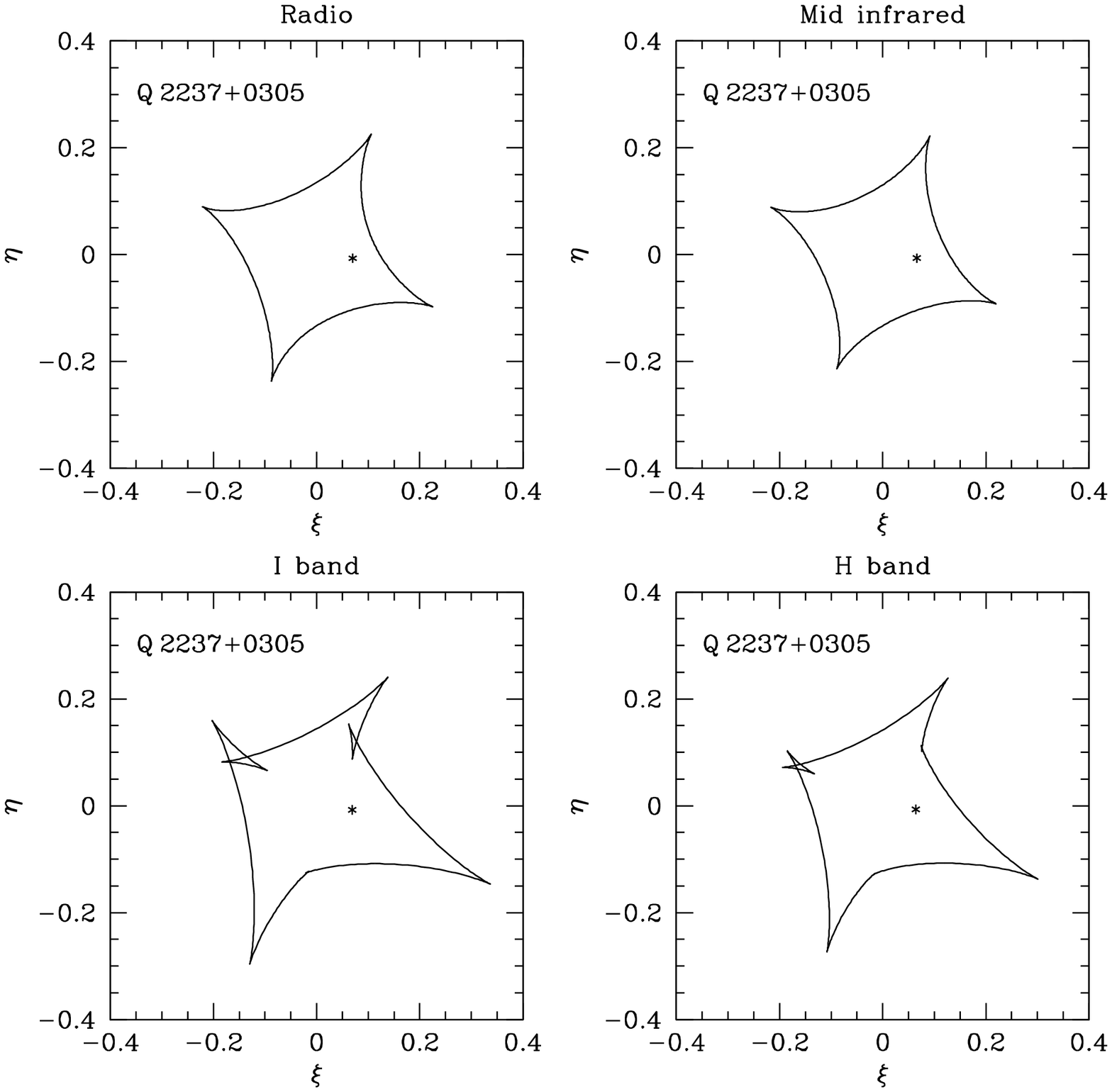}}
\caption{The panels show the caustics of the Einstein Cross
corresponding to the exact solutions (dashed curves in Fig. 2).  The
computed position of the lensed quasar is indicated. It always lies in
the region of 4 images. For the I and H band, there are butterfly and
swallowtail cusps (small regions of the source plane where 6 images
may occur).}
\label{fig:caustic}
\end{figure*}

\section{Fitting Gravitational Lenses}

\subsection{A Singular Value Decomposition Problem}

First, we insert the Fourier series of $F(\theta)$ into
eqs.~(\ref{eq:xi})--(\ref{eq:eta}) to establish the linear equations
for fitting for $\beta = 1$.  Introducing a useful notation, we write
for the lens equation
\begin{eqnarray}
\xi(r, \theta) &=& r \cos\theta\!-\!{ a_0 \over 2 } \alpha_0(\theta)
\!-\!\sum_{k = 1}^{\infty}
[a_k \alpha_k (\theta)\!+\!b_k \beta_k (\theta)], \\
\eta(r, \theta) &=& r \sin\theta\!-\!{ a_0 \over 2 }\hat\alpha_0(\theta) 
\!-\!\sum_{k = 1}^{\infty}
[a_k \hat\alpha_k(\theta)\!+\! b_k \hat\beta_k(\theta)],
\end{eqnarray}
where $a_k$ and $b_k$ are the Fourier coefficients of the unknown
function $F(\theta)$, which describes the angular part of the
potential.  For the coefficients $\alpha_k$ and $\beta_k$, we find
\begin{eqnarray}
\alpha_k(\theta) &=& \cos\theta \cos(k\theta) + k\sin\theta
\sin(k\theta), \nonumber \\ 
\beta_k(\theta) &=& \cos\theta \sin(k\theta) - k \sin\theta \cos(k\theta),
\end{eqnarray}
and for $\hat\alpha_k$ and $\hat\beta_k$, we find
\begin{eqnarray}
\hat\alpha_k(\theta) &=&  \sin\theta \cos(k\theta) - k \cos\theta
\sin(k \theta),
\nonumber \\
\hat\beta_k(\theta) &=&  \sin\theta \sin(k\theta) + k \cos\theta\cos(k\theta),
\end{eqnarray} 
for all $k \geq 0$. The lens equations~(\ref{eq:xi})--(\ref{eq:eta})
must hold at the positions of the $n$ images, given in polar
coordinates as ($ r_\ell, \theta_\ell$) with $\ell = 1, ... , n$. This
provides $2n$ constraints on the unknown source position $(\xi, \eta)$
and the unknown Fourier coefficients.

Now let us consider the magnification of the $\ell$th image. For the
flat rotation curve case ($\beta = 1$), we obtain
\begin{equation}
\det J = {1 \over p_\ell \mu_\ell} = 1 - {G(\theta_\ell) \over r_\ell},
\label{eq:cc}
\end{equation}
where $p_\ell$ is the parity of the image $\ell$. It is only the flux
ratios that can be related to the observational data, not the
magnifications themselves.  For the flux ratio of image $\ell$
compared to image $k$, we obtain
\begin{equation}
f_{k\ell} = { p_k \mu_k \over p_\ell \mu_\ell } = { 1 - G(\theta_\ell) / r_\ell \over
1- G(\theta_k) / r_k}.
\end{equation}
The flux ratio may be positive or negative depending on the
combination of parities of the images.  For flux ratios, we always
compare to a reference image, which without loss of generality we take
to be the first image. As we take ratios, we only need to identify
images of the same parity, which in a quadruplet lens are usually on
opposite sides of the lensing galaxy. Clearing the fractions, we can
write the equation in a similar form to the lens equation
\begin{equation}
(f_{1\ell}\!-\!1) r_1 r_\ell  = { a_0 \over 2 } \gamma_0(\theta_\ell)
\!+\!\sum_{k  = 1}^{\infty}
[a_k \gamma_k (\theta_\ell)\!+\!b_k \delta_k (\theta_\ell)],
\end{equation}
with
\begin{eqnarray}
\gamma_k (\theta_\ell) &=& (1\!-\!k^2)
[f_{1\ell} r_\ell \cos(k \theta_1)\!-\!r_1 \cos(k \theta_\ell)]
\nonumber \\
\delta_k (\theta_\ell) &=&  (1\!-\!k^2)
[f_{1\ell} r_\ell \sin(k \theta_1)\!-\!r_1 \sin(k \theta_\ell)].
\end{eqnarray}
This provides a further $n\!-\!1$ linear constraints on the unknowns.

Now let us introduce a compact notation to facilitate matters. We
set $\alpha_{k\ell} \equiv \alpha_k (\theta_\ell)$, where the first
index denotes the Fourier component and the second index the image
number. Exactly similar definitions can be made for $\beta_{k\ell},
\hat\alpha_{k\ell},\hat \beta_{k \ell}, \gamma_{k\ell}$ and
$\delta_{k\ell}$.  We are now ready to set up the matrix equation for
the fitting to the positions of the images and the flux ratios. Let us
write
\begin{equation}
C \bx = \bd
\label{eq:matrix}
\end{equation}
where
\begin{equation}
{\bd} = (r_\ell \cos\theta_\ell, ..., r_\ell \sin\theta_\ell, ..., 
(f_{1\ell} - 1)r_1 r_\ell, ... )^T
\label{eq:observables}
\end{equation}
is a vector of $3n\!-\!1$ observables. Here, $\ell$ runs from
$1$ to $n$ for a lens system with $n$ images.  There is always one
less equation for the flux constraints than for the image position
constraints. The vector
\begin{equation}
{\bx} = (\xi, \eta, a_0 / 2, a_1, b_1, a_2, b_2, ....)^T
\label{eq:vectorx}
\end{equation}
contains the unknown quantities, namely the source coordinates ($\xi,
\eta$), and the Fourier coefficients. There are in principle
infinitely many such coefficients, but in practice the Fourier series
is usually terminated so that ${\bx}$ is a vector of $3n\!-\!1$
components as well.  Finally, for the matrix $C$, we obtain
\begin{equation}
C = \left( \matrix{
1 & 0 & \alpha_{0\ell} & \alpha_{1\ell} & \beta_{1\ell} & \alpha_{2\ell} 
& \beta_{2\ell} & ... \cr
. & . &  .          & .           &  .         & .           &  . & ... \cr 
0 & 1 &\hat\alpha_{0\ell} & \hat\alpha_{1\ell} & \hat\beta_{1\ell} &
\hat\alpha_{2\ell} & \hat\beta_{2\ell} & ... \cr
. & . &  .          & .           &  .         & .           &  . &
... \cr 
0 & 0 & \gamma_{0\ell} & \gamma_{1\ell} & \delta_{1\ell} &
\gamma_{2\ell} & \delta_{2\ell} & ... \cr
. & . &  .          & .           &  .         & .           &  . &
... \cr 
} \right)
\label{eq:defC}
\end{equation}
Provided we can measure all $n$ image positions and $n-1$ flux ratios,
then $C$ has $3n-1$ rows. The number of columns is in principle
arbitrarily large, although in practice it usually makes sense to
terminate the Fourier series so that $C$ is a $3n\!-\!1 \times
3n\!-\!1$ matrix.

As it stands, the matrix $C$ is always singular $(\det C = 0)$. There
exists a null space for the matrix $C$, i.e., there exists at least
one non-vanishing vector ${\bx_0}$ which satisfies the equation $C
{\bx_0} = {\bf 0}$.  In fact, the matrix $C$ has (at least) a
two-dimensional null space. Fortunately, the null subspace can be easily
constructed.  Since $\alpha_{1\ell} = 1$, $\beta_{1\ell} = 0$,
$\hat{\alpha}_{1\ell} = 0$, $\hat{\beta}_{1\ell} = 1$, $\gamma_{1\ell}
= 0$ and $\delta_{1\ell} = 0$ for all $\ell$, it is easy to verify
that the null vector must be of the form ${\bx}_0 = \lambda_1 {\bv}_1 
+ \lambda_2 {\bv}_2 $ with ${\bv}_1 = (1,0,0,-1,0,...)^T$ and
${\bv}_2 = (0,1,0,0,-1,0,...)^T$.

From the physical point of view, the Fourier coefficients $a_1$ and
$b_1$ simply produce a shift or transformation of the source position
$(\xi,\eta)$.  However, both coefficients do not contribute to the
surface mass density for the special case of $\beta=1$ since
$a_1\cos\theta$ and $b_1\sin\theta$ are solutions of the homogenous
differential equation $F(\theta) + F^{\prime\prime}(\theta) = 0$
(cf. eq.(\ref{eq:ode})). We can excise this degeneracy by simply
setting $a_1 = b_1 = 0$, which corresponds to the choice of origin.
Now we can just remove the 4th and 5th column of the matrix $C$ and
add two higher degree Fourier coefficients to the problem to maintain
$C$ as a $3n\!-\!1 \times 3n\!-\!1$ matrix.  The new matrix is usually
non-singular and so a Gauss-Jordan elimination can formally solve the
problem. However, it is better to use singular value decomposition
(SVD) to solve for the unknown vector $\bx$ (cf. Press et al. 1999).
In this way, we are always guaranteed to find a numerically stable
solution.

The singular value decomposition of the matrix $C$ is
\begin{equation}
C = U \cdot W \cdot V^T,
\end{equation}
where the $3n\!-\!1 \times 3n\!-\!1$ matrices $U$ and $V$ are
orthonormal, and the matrix $W$ is a diagonal matrix of singular
values ($w_1, w_2, ..., w_{3n-1}$).  As usual in a singular value
decomposition (SVD), any very small singular values are removed (see
Press et al. 1989) and this must be accompanied by a corresponding
reduction in the number of unknown Fourier coefficients. Typically,
any singular value whose ratio to the largest singular value is less
than $10^{-4}$ is removed.  After deletion, let us suppose that $m$
singular values remain, where $m \le 3n-1$. If $m = 3n-1$, then the
SVD solution is exact; if singular values are removed, then the
solution is usually very good.

Let us remark here that one can add further higher degree Fourier
coefficients and then the fitting problem becomes underdetermined.  We
can obtain a whole space of models all of which would equally well
reproduce the observed data. It is clear -- even with the severe
restriction that we have made to a scale-free model with a flat
rotation curve -- that the variety of models that fit the data is
extremely large. This cautions us against making rash statements based
on goodness-of-fit to a single model.

\subsection{Uncertainties}

It is of modest interest to provide a model that fits the
observational data exactly, as the uncertainties in the flux ratios
may be substantial. It is of much greater interest to be able to find
a set of models that can reproduce the data within the errors.

The relative image positions are usually known to extremely high
accuracy. For lenses observed with the {\it Hubble Space Telescope},
the error in the relative astrometry is only $\sim 3$ mas.  From the
point of view of modelling, however, the important quantities are the
relative positions of the images with respect to the centre of the
mass distribution of the lensing galaxy. These coordinates are less
accurately determined. The centre of the light distribution of the
lensing galaxy is sometimes poorly known (especially if it is at high
redshift). In any case, there is no guarantee that the center of the
light corresponds to the centre of the mass. In our modelling, we
always reproduce the relative image positions exactly, but we allow
for an error of no more than 50 mas in the position of the center of
the lens.

For the flux ratios, we distinguish between the radio and the optical
data. The flux measurements in the radio may be affected by
scintillation or scatter-broadening or free-free absorption. An
isolated example of radio microlensing is known (Koopmans \& de Bruyn
2000), but the effects of microlensing are not generally
believed to be substantial in the radio.  Whenever we use radio data,
we assume that the typical uncertainty in the flux ratio is no more
than 5 per cent.  However, there are much greater problems in the optical
bands. Each image might be affected by differential extinction or by
microlensing. Microlensing can cause an image to be below or above its
average magnification for decades (see Witt \& Mao 1994 for an
extensive discussion of the effects of microlensing). The recent years
have seen monitoring campaigns of both Q\,2237+030 and Q\,0957+561,
which provide irrefutable evidence for microlensing (Colley et
al. 2000; Wozniak et al. 2002; Schmidt et al. 2002). There are at
least a further five multiply imaged quasars for which evidence exists
implicating microlensing in the optical (HE\,1104-1805, PG\,1115+080,
H\,1413+117, B\,0218+357, B\,1600+434). Therefore, the optical flux
ratios are more uncertain and we assume that they have an error of
$\lta 10$ per cent.

\begin{table*}
\begin{center}
\begin{tabular}{lccccccc} \hline
\null & $-\Delta \alpha$ & $\Delta \delta$  & I band (K93) & H band & V band
& I band \\ 
\null & (in ${}^{\prime\prime}$) 
& (in ${}^{\prime\prime}$) & (in mags) & (in mags) & (in mags) & (in
mags) \\ \hline
A1 & $-1.328\pm 0.004$ & $-2.034 \pm 0.004$ & $16.12$  
& $15.71 \pm 0.03$ & $16.90 \pm 0.11$ & $16.42 \pm 0.02$ \\ 
A2 & $-1.477\pm 0.004$  & $-1.576 \pm 0.003$ & $16.51$ 
& $16.21 \pm 0.04$ & $17.62 \pm 0.09$ & $16.85 \pm 0.03$ \\ 
B & $0.341\pm 0.003$ & $-1.961 \pm 0.004$ & $18.08$ 
& $17.70 \pm 0.05$ & $18.39 \pm 0.45$ & $17.91 \pm 0.39$ \\ 
C & $0$  & $0$ & $17.58$ & $17.23\pm 0.04$ & $18.95 \pm 0.32$ 
& $18.37 \pm 0.34$ \\ 
G & $-0.381 \pm 0.003$ & $-1.344 \pm 0.003$ & \null & \null & \null &
\null \\ \hline
\end{tabular}
\end{center}
\caption{Observational data on the quadruplet PG\,1115+080. The
optical positions are taken from the CASTLES survey, as are the V,H
and I band fluxes.  For comparison, we also give the earlier I band
fluxes found by Kristian et al. (1993).}
\label{table:secondlens}
\end{table*}
\begin{table*}
\begin{center}
\begin{tabular}{ccccccccccc} \hline
Band & $\xi$ & $\eta$ & $a_0$ & $a_2$ & $b_2$ & $a_3$ & $b_3$ & $a_4$ 
& $b_4$ \\ \hline
H& $-0.0273$ & 0.1122 & 2.2946 & $-0.0029$ & 0.0052 & $-0.0021$ & 0.0027 &
$-0.0007$ & $-0.0002$ \\ 
I (K93)&  $-0.0277$ & 0.1114 & 2.3039 & $0.0007$ & 0.0059 & 0.0013 & 0.0060
& 0.0000 & 0.0020 \\ 
V&  $-0.0238$ & 0.1175 & 2.2871 & $-0.0081$ & 0.0035 & $-0.0054$ & $-0.0001$ &
$-0.0022$ & $-0.0020$ \\
I&  $-0.0218$ & 0.1145 & 2.2880 & $-0.0064$ & 0.0034 & $-0.0049$ & $0.0001$ &
$-0.0016$ & $-0.0012$ \\
\hline
\end{tabular}
\end{center}
\caption{Fourier coefficients of the solutions for PG\,1115+080,
assuming a constant external shear with a magnitude of $0.103$ and a
direction of $65.8^\circ$ measured from west to north (following
Schechter et al. 1997).  The coefficients are computed in the same
coordinate system as the data in Table~\ref{table:secondlens} but
translated to the lens centre.}
\label{table:modelstwo}
\end{table*}
\begin{table*}
\begin{center}
\begin{tabular}{lcc|cc|cc} \hline
\null & \multicolumn{2}{c}{$-\Delta \alpha$ (in ${}^{\prime\prime}$)}& 
 \multicolumn{2}{c}{$\Delta \delta$ (in ${}^{\prime\prime}$) } & 
  \multicolumn{2}{c}{H band fluxes (in mags)} \\
\null & Predicted & Observed & Predicted & Observed 
& Predicted & Observed \\ \hline
A1 & $-1.324$ & $-1.328$ & $-2.035$ & $-2.034$ 
& $15.71$ & $ 15.71$  \\
A2 & $-1.481$ & $-1.477$ & $-1.575$ & $-1.576$ 
& $16.10$ & $16.21$ \\
B & $0.341$  & $0.341$  & $-1.961$ & $-1.961$ 
& $17.59$ & $17.70$\\
C & $0$  & $0$  & $0$  & $0$  & $17.33$ & $17.23$  \\
G & $-0.391$ & $(-0.381)$ & $-1.334$ & $(-1.344)$ & $-$& $-$\\ \hline
\null & \multicolumn{2}{c}{$-\Delta \alpha$ (in ${}^{\prime\prime}$)}& 
 \multicolumn{2}{c}{$\Delta \delta$ (in ${}^{\prime\prime}$) } & 
  \multicolumn{2}{c}{I band fluxes of K93 (in mags)} \\
\null & Predicted & Observed & Predicted & Observed 
& Predicted & Observed \\ \hline
A1 & $-1.324$ & $-1.328$ & $-2.035$ & $-2.034$ 
& $16.12$ & $ 16.12$  \\
A2 & $-1.481$ & $-1.477$ & $-1.575$ & $-1.576$ 
& $16.40$ & $16.51$ \\
B & $0.341$  & $0.341$  & $-1.961$ & $-1.961$ 
& $18.18$ & $18.08$\\
C & $0$  & $0$  & $0$  & $0$  & $17.47$ & $17.58$  \\
G & $-0.386$ & $(-0.381)$ & $-1.329$ & $(-1.344)$ & $-$& $-$\\ \hline
\end{tabular}
\end{center}
\caption{Predicted and observed quantities for two of the solutions
given in Figure~\ref{fig:secondlens}. The upper (or lower) panel
corresponds to the lens model with equidensity contour given by the
bold curve in the panel for the H band (or I band of K93) fluxes.  The
positions $-\Delta \alpha$ and $\Delta \delta$ are computed relative
to image C.  Notice that astrometric data are well-reproduced within
even the formal uncertainties whilst the flux ratios are not
anomalous, once an uncertainty of 10 per cent is allowed for the
effects of microlensing.}
\label{table:secondlensresults}
\end{table*}
\begin{figure*}
 \epsfysize=12.cm \centerline{\epsfbox{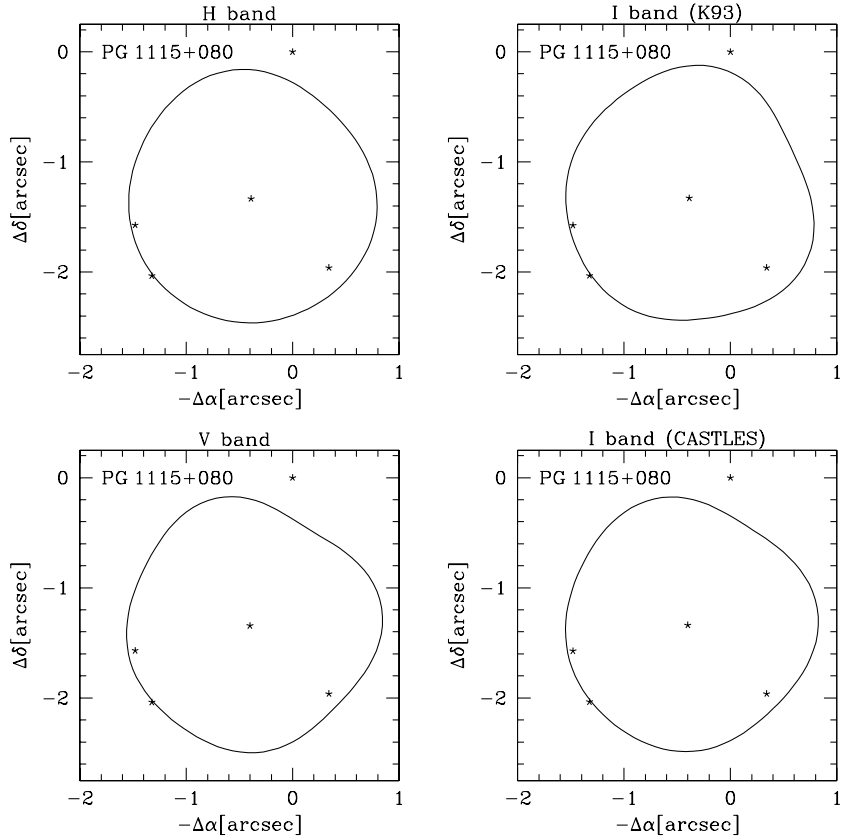}}
\caption{The panels show the equidensity contours ($\kappa = 1/2$) for
lens models of PG\,1115+080. The bold curves reproduce the relative
astrometry within the errors and the measured optical flux ratios to
better than 10 per cent. As a constant external shear is assumed, the
equidensity contour is no longer a critical curve. The locations of
the four images and the lensing galaxy are also marked.}
\label{fig:secondlens}
\end{figure*} 
\begin{figure*}
 \epsfysize=12.cm \centerline{\epsfbox{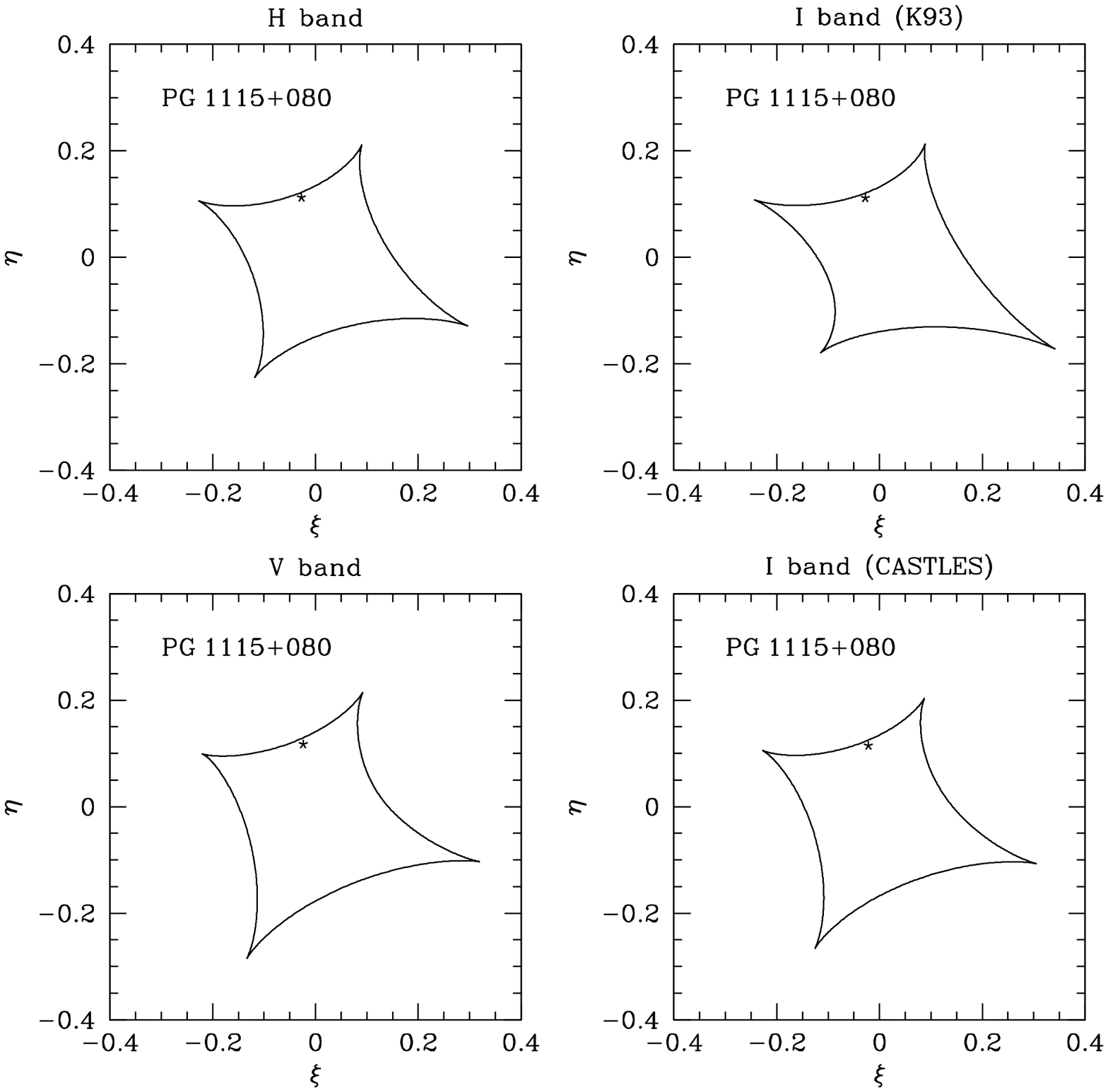}}
\caption{The panels show the caustics of the lens model for
PG\,1115+080 corresponding to the solutions in Fig. 4.  The computed
position of the lensed quasar is indicated. It always lies in the
region of 4 images. There are no regions in which higher order imaging
occurs (butterfly and swallowtail cusps).}
\label{fig:caustic1115}
\end{figure*}

\subsection{Position Angle of the Lensing Galaxy Axes}
 
The axes of the local right ascension and declination coordinates may
be rotated by an arbitrary angle with respect to the major and minor
axes of the lensing galaxy. This is implicitly contained in the
derived Fourier coefficients.

We can expand the Fourier series as
\begin{eqnarray}
& &\sum_{k=1}^\infty [a_k \cos(k\theta) + b_k \sin(k \theta)] \nonumber \\
&=& \sum_{k=1}^\infty [a_k^R \cos(k\theta + k\theta_R) + b_k^R
\sin(k\theta +k\theta_R)],
\end{eqnarray}
where $\theta_R$ is the rotation angle and $a_k^R$ and $b_k^R$ are the
Fourier coefficients in the rotated coordinate system.  It is easy to
verify that
\begin{eqnarray}
a_k^R & = & a_k \cos k\theta_R\!+\!b_k \sin k\theta_R \nonumber \\ 
b_k^R & = & -a_k\sin k\theta_R\!+\!b_k \cos k\theta_R.
\end{eqnarray}
If the lensing galaxy has a major axis, the following condition must
be satisfied:
\begin{equation}
\label{eq:thetaR}
 G^\prime(\theta_R) \approx 0 \approx 
G^\prime(\theta_R + 180^\circ).  
\end{equation}
If the lensing galaxy looks regular, it probably has, in addition to a
major axis, an approximate reflection symmetry $G(\theta) \approx
G(-\theta)$, or equivalently $F(\theta) \approx F(-\theta)$ .
Therefore, in the rotated system, we expect $b_k^R \approx 0$ for all
$k$.  In other words, the rotated system is defined by
\begin{equation}
\sum_{k = 1}^n (b_k^R)^2 = {\rm minimum}.
\end{equation}
Solving this equation yields the rotation angle of the major axis
of the lensing galaxy
\begin{equation}
\label{eq:thetaRmin}
2 \sum_{k = 1}^n k a_k b_k \cos (2k \theta_R) =  \sum_{k = 1}^n k(a_k^2 -
b_k^2) \sin (2k \theta_R).
\end{equation}
This equation must in general be solved numerically. For galaxy-like
mass distributions, the dominant higher-order terms in the Fourier
expansion are those with $k=2$ and then the solution is analytic,
namely
\begin{equation}
\tan 2\theta_R  = {b_2 \over a_2}.
\end{equation}
For applications to individual lenses, we note that the position angle
as conventionally defined by astronomers (North through East) is
$90^\circ - \theta_R$.

\subsection{Mass inside the Einstein Ring}

Evans \& Witt (2001) showed that the mass inside the Einstein ring
must be very close to the mass inside the critical curve for the
special case of $\beta = 1$, and can be written as
\begin{equation}
M_E \approx M_{\rm crit. curve} = \frac{1}{2\pi} \int_0^{2\pi}
G^2(\theta) d\theta.
\end{equation}
We therefore deduce that
\begin{equation}
M_E \approx {a_0^2 \over 4} + \sum_{i=2}^N (k^2 - 1)^2 (a_k^2 +
b_k^2).
\end{equation}
To apply this to a real lens, we have to scale the result by the
factor $\Sigma_{\rm crit} \Dd^2$ containing the normalized surface mass
density which is is given by
\begin{equation}
\Sigma_{\rm crit} = \frac{c^2 \Ds}{4\pi G \Dd \Dds}.
\label{eq:critdens}
\end{equation}
Here, $\Dd, \Ds$ and $\Dds$ are the distance to the deflector, the
distance to the source and the distance from the deflector to the
source respectively.  Given the redshift of the lens and the source,
this (\ref{eq:critdens}) enables the projected mass within the
Einstein ring to be estimated.

\section{AN APPLICATION: ANOMALOUS FLUX RATIOS}

In this section, we consider three of the lens systems for which
substructure has been claimed either by Metcalf \& Zhao (2002) or by
Dalal \& Kochanek (2002) or by Chiba (2002).

\subsection{Q\,2237+030 (The Einstein Cross)}

The data on the four images of Q\,2237+030 (the Einstein Cross) are
listed in Table~\ref{table:ecross}. Much of this comes from the
CASTLES survey~\footnote{http://cfa-www.harvard.edu/castles/}.
Q\,2237+030 is an unusual lens because so much more information is
available for the lensing galaxy than is customary. In particular, the
redshift of the lens is so small ($z =0.039$) that the galaxy's light
distribution can be measured (e.g., Wyithe et al.  2002). The lens is
a face-on barred galaxy (e.g., Schmidt, Webster \& Lewis 1998). The
optical fluxes are known to be affected by microlensing (Irwin et
al. 1989; Wambsganss, Schneider \& Paczy\'nski 1990; Wozniak et
al. 2000), which causes each image to differ from the average
magnification.  Typically, we expect that the images are very likely
below the average magnification if they are in a quiescent state (for
years), while the images are very likely above the average
magnification if they are in a more active state (cf. Witt \& Mao
1994).
 
It is often argued that the radio and mid-infrared fluxes are more
reliable because the source emitting region is more extended and
therefore less affected by microlensing.  If this is the case, then
gravitational lens models should use the radio and mid-infrared flux
ratios in preference to optical. The flux ratios in the radio were
measured by Falco et al. (1996) with a signal-to-noise of $\sim
2-4$. The mid-infrared flux ratios have recently become available
thanks to Agol, Jones \& Blaes (2000). The flux ratios in radio and
mid-infrared are in good agreement, but differ from the optical flux
ratios.

Figure~\ref{fig:ecross} shows the critical curves for four models of
Q\,2237+030. From eq.~(\ref{eq:cc}), the critical curve is also an
equidensity contour when the rotation curve is flat. As the models are
scale-free, the equidensity contours retain the same shape independent
of position, although the size of course varies.  In each case, the
dashed curve describes a model that exactly reproduces the image
positions and flux ratios. It is found by solution of the matrix
equation~(\ref{eq:matrix}) using singular value decomposition (SVD).
Figure~\ref{fig:caustic} shows the corresponding caustics of the exact
solutions of the critical curves (dashed lines).  The more irregular
critical curves (I and H band data) correspond to caustics with
over-folds like swallowtails and butterflies.  However, in each case,
the source position still lies inside the four image region. Unwanted
extra images are not produced.

We assume that the measured flux ratios in the optical band are in
error by $\lta 10$ per cent due to microlensing. The relative
positions of the image are always maintained within the formal errors
(given in Table~\ref{table:ecross}), but the center of the lensing
galaxy may be displaced by up to 50 mas. We find the model for which
\begin{equation}
b_2^2 + \sum_{i \ge 3} [a_i^2 + b_i^2]
\label{eq:minimum}
\end{equation}
is minimised. This suppresses Fourier components higher than the
bar-like $m=2$ component.  In other words, out of all the solutions
that reproduce the data, we choose the one that most looks like a
galaxy.  The models so produced are shown in bold lines in
Figure~\ref{fig:ecross}; the details of the solutions are listed in
Table~\ref{table:modelsone}.  Finally, for two of the bold line
solutions, we show how the image positions and the flux ratios are
reproduced in Table~\ref{table:ecrossresults}. All the observable
quantities are extremely well reproduced by the model, including the
flux ratios.

Two points emerge clearly from this. First, the equidensity contours
implied by reproducing the exact radio or mid-infrared fluxes are
already in good agreement. The position angle of the lensing galaxy as
inferred from eq.~(\ref{eq:thetaR}) using the radio or mid-infrared
solutions is $\sim 67.3^\circ$ (measured from North through East).
For comparison, Trott \& Webster (2002) report that the position angle
of the bar in the lensing galaxy is $\sim 39^\circ$, while the major
axis of the galaxy in its outer parts is $\sim 77^\circ$.  Second, the
equidensity contours implied by exactly reproducing the I and the H
band optical flux ratios are distorted.  However, there are more
regular solutions once an allowance is made for the effects of
microlensing on the flux ratios.

Distortions of isophotes from elliptical form are generally given in
terms of Bender et al.'s (1989) coefficients. These are usually
calculated for early-type galaxies and have absolute values less than
0.05 (see e.g., Binney \& Merrifield 1998, section 4.3).  In fact,
barred galaxies, such as the lens in Q\,2237+030, may have
substantially larger deviations, which moreover depend on position
along the major axis. The method of calculating these coefficients for
densities given as Fourier expansions is elaborated in Appendix C; we
quote the results here.  For the radio contours, Bender's coefficients
are $a_3^B/a_0 = -0.008$ and $a_4^B/a_0 = -0.005$; for the mid-infrared,
they are $a_3^B/a_0 = 0.001$ and $a_4^B/a_0 = -0.011$. These values are
well within the observed range, emphasising that our solutions for the
shapes of the equidensity contours are realistic. By comparison, for
the distorted I band contours, the coefficients are $a_3^B/a_0 = 0.012$
and $a_4^B/a_0 = -0.034$.

All this provides strong evidence for the point of view advanced in
Agol et al. (2000) that the radio and mid-infrared fluxes are
trustworthy and that the optical flux ratios are discrepant because of
microlensing.  All the data are internally consistent and once
reasonable error bars are placed on the flux ratios, there is no
need to invoke substructure to explain the data (c.f. Metcalf \& Zhao
2002).

\begin{table*}
\begin{center}
\begin{tabular}{lccccccc} \hline
\null & $-\Delta \alpha$ & $\Delta \delta$ & Radio fluxes 
& r band & V band & H band \\ 
\null & (in ${}^{\prime\prime}$)
& (in ${}^{\prime\prime}$) & (in $\mu$ Jy) & (in mags) & (in mags)
& (in mags) \\ \hline
A & $0.0$             & $0.0$             & $216$  
& $16.77$ & $16.43 \pm 0.11$ & $14.41 \pm 0.03$ \\ 
B & $0.385\pm 0.003$  & $-0.317 \pm 0.003$ & $221$ 
& $16.45$ & $16.45 \pm 0.10$ & $14.29 \pm 0.03$\\ 
C & $0.722\pm 0.003$ & $-1.068 \pm 0.003$ & $115$  
& $17.25$ & $17.09 \pm 0.07$ & $14.98 \pm 0.03$\\ 
D & $-0.562\pm 0.004$  & $-1.120 \pm 0.003$ & $4.5$ 
& $20.40$& $20.44\pm 0.06$ & $18.14 \pm 0.04$ \\ 
G & $-0.375 \pm 0.004$ & $-0.973 \pm 0.004$ & \null & $\null$ \\ \hline
\end{tabular}
\end{center}
\caption{Observational data on the quadruplet B\,1422+231. The optical
positions are taken from the CASTLES survey, as are the H and V band
fluxes.  The radio fluxes are provided by Patnaik et al. (1992).}
\label{table:thirdlens}
\end{table*}
\begin{figure*}
 \epsfysize=12.cm \centerline{\epsfbox{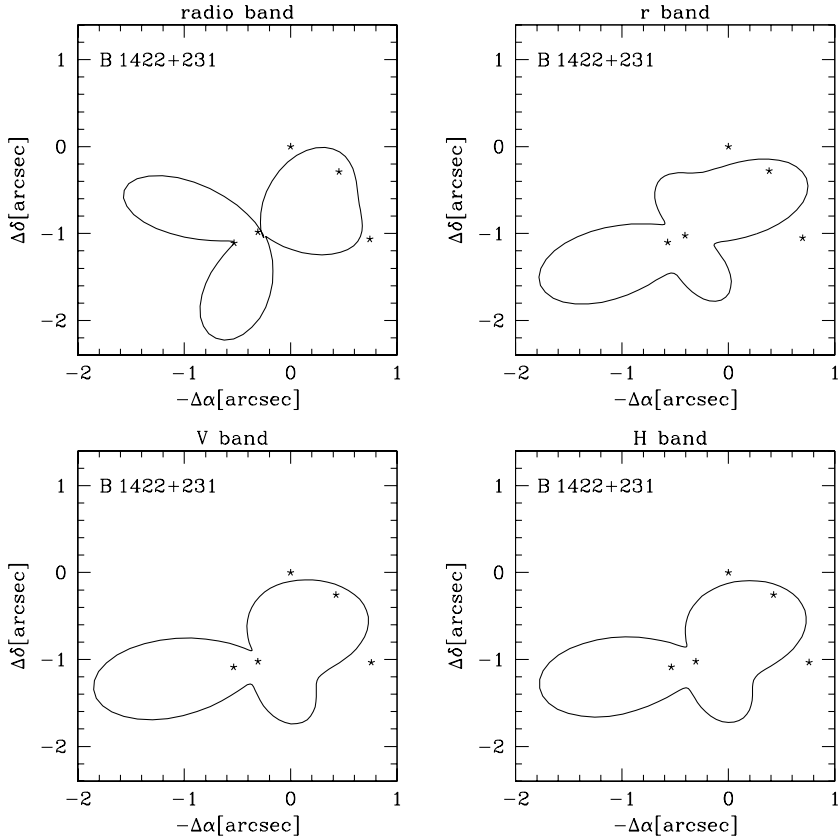}}
\caption{The panels show the equidensity contours (or critical curves)
for the quadruplet system B\,1422+231. Working clockwise from the top
left, the panels use the flux ratios in the H band, r band, V band and
radio. This is a lens for which substructure may well be present, as
none of our models are at all satisfactory.}
\label{fig:thirdlens}
\end{figure*}

\subsection{PG\,1115+080}

Let us now examine the lens system PG\,1115+080, which has been
claimed to provide evidence for substructure by Dalal \& Kochanek
(2002), Metcalf \& Zhao (2002) and Chiba (2002).  The observational
data on PG\,1115+080 are listed in Table~\ref{table:secondlens}.  The
I, V and H band fluxes are provided from the CASTLES survey. There is
also earlier data on the I band fluxes thanks to Kristian et
al. (1993, hereafter K93).

There is already evidence that the flux ratios may come with
substantial uncertainties.  For example, Foy, Bonneau \& Blazit (1985)
found component A2 of PG\,1115+080 was brighter than A1 in 1984
March-May, whereas most subsequent investigators have found A2 to be
fainter than A1. The conclusion advanced by Christian, Crabtree \&
Waddell (1987) is that the images in PG\,1115+080 are being affected
either by microlensing or by time-delayed intrinsic variations in the
quasar or both.  Witt, Mao \& Schechter (1995) showed that the
likelihood of observing microlensing in images A1 and A2 is much
higher than in C and D.  Notice too that the CASTLES and the Kristian
et al (1993) I band fluxes in Table~\ref{table:secondlens} are not
consistent within the stated error bars, the discrepancies almost
certainly being due to microlensing.  Radio emission would be less
affected by microlensing, but PG\,1115+080 is radio-quiet and so the
optical fluxes have to be used.

Impey et al. (1998) have carried out deep {\it Hubble Space Telescope}
imaging of PG\,1115+080 in the infrared. They found that the primary
lens is a nearly round elliptical galaxy with an ellipticity of less
than 0.07. The lens is part of a group of galaxies, whose effects must
be included in successful models. The group is modelled by a uniform
external shear with a magnitude of $0.103$ and a direction of
$65.8^\circ$ measured from west to north, following Schechter et
al. (1997).  The extension of our SVD method to the case of uniform
shear is sketched in Appendix D.  We implement our fitting procedure,
deleting two singular values for numerical stability and finding the
solution that minimises (\ref{eq:minimum}), but still reproduces the
data within our assumed errors.  The panels of
Figure~\ref{fig:secondlens} show the equidensity contous of the
resulting models. Again, the relative astrometry of the images is
recovered to within the formal error ($\sim 4$ mas), the lens centre
may be misligned by up to 50 mas, while the optical flux ratios are
reproduced to within 10 per cent.  The comparison between model
predictions and observed quantities for the H and I band (K93)
solutions are given in Table~\ref{table:secondlensresults}.  The
contours implied by reproducing the H or I band (K93) fluxes are in
good agreement and look plausibly like the projected mass distribution
of a very round early-type galaxy, in accordance with Impey et al.'s
(1998) imaging. There is some slight distortion in the equidensity
contours that reproduce the V or I band (CASTLES) flux ratios.  The
caustic networks are shown in Figure~\ref{fig:caustic1115} to
demonstrate that unwanted extra images are not formed.

Again, the shapes of the equidensity contours in
Figure~\ref{fig:secondlens} do exhibit small deviations from pure
elliptical form, which can be quantified by Bender et al.'s (1989)
coefficients. Using the method of Appendix C, we find $a^B_3/a_0 =
0.009$ and $a^B_4/a_0 = -0.004$ for the H band and $a^B_3/a_0 =
-0.015$ and $a^B_4/a_0 = -0.004$ for the I band (K93) contours.  There
is some data on the shapes of dark haloes of spiral galaxies from
numerical simulations (e.g., Heyl, Hernquist \& Spergel 1994), which
supports the notion that $|a_3| \lta 0.05$ and $|a_4| \lta 0.05$. So,
the amplitudes of the higher order terms in our solutions for the lens
galaxy do lie easily within the range expected for dark haloes of
spiral galaxies.

It is clear from the smooth models that we have constructed that there
is no really compelling evidence for substructure in this lens at the
moment, despite abundant claims to the contrary (Dalal \& Kochanek
2002; Metcalf \& Zhao 2002; Chiba 2002; Kochanek \& Dalal 2003).
There are smooth and realistic models that do fit the data within the
likely uncertainties.

\subsection{B\,1422+231}

Patnaik et al. (1992) discovered the gravitational lens B\,1422+231 in
a radio survey. Table~\ref{table:thirdlens} lists some of the
available data on the images, largely provided by the CASTLES group.
It was realised by both Hogg \& Blandford (1994) and Kormann,
Schneider \& Bartelmann (1994) that it is difficult to reproduce the
flux ratios of this lens system with a simple model.  Hogg \&
Blandford used a mixture of isothermal spheres and point masses
whereas Kormann et al. used a model with external shear. The latter
strategy has been followed by most modern modellers.

Using our SVD method, it was impossible to reproduce the reported
errors in the relative astrometry, together with the assumed
uncertainties in the flux ratios (5 per cent for the radio, 10 per
cent for the optical) at least with physical models. Unphysical models
are readily identified as the equidensity contours are
self-intersecting. Accordingly, for this lens only, the errors in the
relative astrometry were inflated by a factor of 10 to $\sim 30$ mas.
Figure~\ref{fig:thirdlens} now shows the results of our fitting
procedure in four flux bands, but none of the results can be said to
be at all satisfactory. The equidensity contours are strongly
distorted from those expected in galaxy-like models. Just as alarming
as the shape of the contours is the fact that the fitted parameters do
not have consistent values from one band to the next.

The conclusion to be drawn by the failure of the fitting procedure is
that -- even taking into account observational uncertainties -- this
lens system requires additional ingredients. Possibly, as Mao \&
Schneider (1998) were the first to suggest, substructure may be
causing a dramatic effect on the flux ratios.  Possibly, as Kundic et
al. (1997) suggest, the effects of external shear caused by nearby
galaxies and clusters of galaxies may also need to be included in the
analysis. 

\section{DISCUSSION}

The main aim of this paper is to present a new, simple approach to
fitting a model to the data on a gravitational lens. Given the image
positions and the flux ratios (if available), a model that exactly
fits that data can be constructed by a simple matrix inversion
(preferably by singular value decomposition, available in standard
numerical libraries). All the fitting parameters enter linearly into
the equation. No complicated non-linear $\chi^2$-fitting needs to be
done.

The lensing galaxy is always assumed to be scale-free with a flat
rotation curve. There is already a substantial body of evidence from
fitting of lenses that early-type galaxies have nearly isothermal
profiles (Kochanek 1995; Mu{\~n}oz et al. 2001; Cohn et al. 2001;
Rusin et al. 2002). In particular, the absence of central images
strongly constrains the lensing potential of early-type galaxies to be
isothermal or steeper and the size of any core region to be small
(e.g., Rusin \& Ma 2001; Evans \& Hunter 2002).  There is both stellar
dynamical (e.g., Gerhard et al. 2001) and X-ray (Fabbiano 1997)
evidence that early-type galaxies have flat rotation curves out to at
least 4 effective radii. Hence, our assumption of a flat rotation
curve seems exactly what the data require.

Unusually, our algorithm allows full flexibility as regards the
angular structure of the lensing potential. Earlier fitting procedures
have allowed the lensing potential to have different radial structure
(for example, different power-law profiles), but usually only simple
forms of angular structure (for example, constant external shear). In
our models, the radial structure is always fixed as isothermal, but
the shape of the density contours is completely arbitrary.  The lens
model has a flexible number of free parameters, and of course includes
the isothermal elliptic density and potential models which have a
distinguished history in gravitational lens modelling.  The fit allows
direct deduction of the mass contour of the lensing galaxy.  Although
we have presented examples based on quadruplet lens systems, the
algorithm can be adapted for a two image or a higher image model.  A
higher image model might be needed, for example, in Q\,0957+561, where
10 sub-image positions can be detected in the radio band. Another good
application might be MG\,0414+0534, where some substructure of the
images are detected (cf. Trotter, Winn \& Hewitt 2000 and their
fitting approach).  Both lenses seem heavily distorted by external
shear. A shear term can be added to our algorithm. If the shear is
known, then the lens and flux ratio equations remain linear.

The flux ratios may enter directly into the fit depending on whether
they are available or trustworthy.  We stress that flux ratios often
do not provide good model constraints.  The reason for this is easy to
understand. The flux ratios depend on the second derivatives of the
lensing potential, whereas time delays and image positions are
proportional to the potential and its first derivative
respectively. Therefore, flux ratios are particularly sensitive to
graininess in the gravitational potential. This often manifests itself
as microlensing. The flux ratios are also affected by the uncertain
differential extinction corrections that must be applied to each
image.  Radio and mid-infrared fluxes are generally more reliable than
optical.  However, effects such as scintillation, scatter-broadening
and free-free absorption may affect the radio fluxes in some of the
lenses. If flux ratios are incorporated into a fit, it is vital that
there is a realistic treatment of the errors.

Our fitting algorithm permits the construction of a whole class of
degenerate models all of which can fit the image position and flux
ratios exactly.  In addition, the models are also degenerate in the
time delay since the delay depends only on the distance of the image
positions to the centre of the lensing galaxy (cf. Witt, Mao \& Keeton
2000).  Such degeneracy is easy to understand pictorially. Given a
Fermat time delay surface, we need only keep the values, the
derivatives and the second derivatives of the surface at the image
positions fixed. We then have enormous freedom to move the surface
(subject only to the constraints that no additional images are
introduced and that no negative mass density is produced).

We have used our new fitting algorithm to examine critically some of
the claims made recently for anomalous flux ratios. The procedure used
by both Dalal \& Kochanek (2002) and Metcalf \& Zhao (2002) was to
show that a family of simple models did not reproduce the flux
ratios. However, it does not follow from this that substructure is
necessary; it may just be that the simple model did not have enough
flexibility to provide a good match. For example, of the five lenses
studied by Metcalf \& Zhao (2002), they themselves concluded that one
(MG\,0414+0534) could be satisfactorily explained by a smooth
model. In this paper, we have demonstrated that the data on two others
(PG\,1115+080, Q\,2237+030) are consistent with smooth galaxy-like
models, especially when a realistic treatment of the errors in the
flux ratios is incorporated.

We note that the fraction of mass in substructure expected in galaxy
haloes is very small ($\lta 5$ per cent) and it occurs overwhelmingly in the
outlying portions (e.g., Moore et al. 1999; Ghigna et al. 2000).
Substructure evolves as it is subjected to tides, impulsive collisions
and dynamical friction. Tidal disruption becomes important as soon as
the mean density of the galaxy is equal to the density of the
substructure. Similarly, the dynamical friction timescale scales with
the square of the galactocentric radius (see e.g., Binney \& Tremaine
1987). So, both tides and dynamical friction are efficient at erasing
substructure in the inner parts. This effect is referred to as
``anti-biasing'' by Ghigna et al. (2000). So, on dynamical grounds,
little substructure is projected within the Einstein radius.  Taking
the example of PG\,1115+080, the projected Einstein radius is just
$\sim 3.6$ kpc. The fraction of mass in substructure projected within
a cylinder of radius $\sim 3.6$ kpc is clearly much lower than the
global fraction of $5$ per cent, which pertains to the entire dark halo of
total extent $\sim 200$ kpc. It is crucial that lensing calculations
do not assume an everywhere uniform fraction of substructure in the
halo, as this does not take into account the ``anti-biased'' spatial
distribution of substructure and therefore necessarily over-emphasises
the importance of the effects of substructure on flux ratios. Notice
that an interesting consequence of the spatial distribution is that
anomalous flux ratios are more likely to occur for lenses with larger
angular separation, as these have larger Einstein radii.

The real question at issue is the following. Suppose a simple lens
model (such as a simple isothermal ellipsoid plus shear) does not
adequately fit the data on positions and flux ratios. What can be
legitimately deduced? The difficulty is that there are many
modifications of the simple model that remove the discrepancy. One of
these is substructure, as pointed out by Dalal \& Kochanek and Metcalf
\& Zhao.  As shown in this paper, another is higher order multipoles
in the lensing mass, such as diskiness, boxiness, lopsidedness and
barredness (see also M\"oller, Hewett \& Blain 2003, who make a
similar point).  It is crucial to develop techniques to distinguish
between higher order quadrupoles on the one hand and substructure on
the other.  From our modelling, we tend to agree that the substructure
candidate B\,1422+231 originally pointed out by Mao \& Schneider
(1998) is strong.  Using a novel application of the cusp relation,
Keeton, Gaudi \& Petters (2002) have argued that B\,2045+265 and RX
J\,0911+0551 may be two further good candidates.

It is an outstanding problem to predict -- for different cosmologies
-- how many quadruplets may have anomalous flux ratios. Accurate
calculations will become possible only when distributions of mass and
position of substructure become available from high resolution
simulations. It will be interesting to see whether the numbers of
lenses for which substructure is currently being claimed are
compatible with cold dark matter theories.

\section*{Acknowledgments}
We acknowledge with gratitude the remarkable service provided to the
community by the CASTLES website (C.S. Kochanek, E.E. Falco, C. Impey,
J. Lehar, B. McLeod, H.-W. Rix). NWE thanks the Royal Society for
financial support and HJW thanks the sub-Department of Theoretical
Physics, Oxford for their hospitality during working visits. NWE
acknowledges Ian Browne, Leon Koopmans and Jerry Ostriker for a number
of interesting discussions. Shude Mao and Olaf Wucknitz are thanked
for a helpful comments on the draft manuscript.

\begin{appendix}

\section{Full Solution by the Variation of Parameters}

Observationally speaking, it is the angular structure of the density
contours $G(\theta)$ that is more accessible. It is sometimes useful
to be able to generate the corresponding angular structure of the
lensing potenial $F(\theta)$. This can be done by solving
eq~(\ref{eq:ode}) using the method of variation of the parameters (e.g.,
Bronshtein \& Semendyayev 1998, section 3.3.1.3.4; Evans \& Witt 2001)
to give
\begin{eqnarray}
F(\theta) &=& 
{ \sin(\beta \theta ) \over \beta }
\Bigl[ C_1 + \int_0^\theta G(\vartheta) \cos(\beta
\vartheta) \,\d\vartheta \Bigr] \nonumber \\
&-&  {\cos(\beta \theta)\over \beta } \Bigl[
C_2 + \int_0^\theta G(\vartheta) 
\sin(\beta\vartheta) \,\d\vartheta \Bigr].
\label{eq:crux}
\end{eqnarray}
This establishes the relation between the potential and the surface
mass density in terms of two constants $C_1$ and $C_2$, whose values
will be given shortly.

Therefore, the deflection angle has components
\begin{eqnarray}
\phi_x &=& r^{\beta\!-\!1}\sin (\beta\!-\!1)\theta 
\Bigl[ C_1 + \int_0^\theta G(\vartheta) \cos(\beta
      \vartheta) \,\d\vartheta \Bigr] \nonumber \\
& & -r^{\beta\!-\!1}\cos (\beta\!-\!1)\theta \Bigl[ C_2 + \int_0^\theta G(\vartheta) 
         \sin(\beta\vartheta) \,\d\vartheta \Bigr], \nonumber \\
\phi_y &=& r^{\beta\!-\!1}\cos (\beta\!-\!1)\theta 
\Bigl[ C_1 + \int_0^\theta G(\vartheta) \cos(\beta
      \vartheta) \,\d\vartheta \Bigr] \nonumber \\
& & +r^{\beta\!-\!1}\sin (\beta\!-\!1)\theta \Bigl[ C_2 + \int_0^\theta G(\vartheta) 
         \sin(\beta\vartheta) \,\d\vartheta \Bigr].
\label{eq:defangle}
\end{eqnarray}
Another expression can be given for the deflection angle, namely

\begin{eqnarray}
\phi_x &=& {\beta x \over r^{2-\beta}}F(\theta) - {y \over
r^{2-\beta}} F^\prime(\theta), \nonumber \\
\phi_y &=& { \beta y \over r^{2-\beta}}F(\theta) + {x \over
r^{2-\beta}} F^\prime(\theta).
\label{eq:defanglenew}
\end{eqnarray}
By substituting the Fourier series into
eqs.~(\ref{eq:defangle})--(\ref{eq:defanglenew}), we obtain
expressions for the constants
\begin{eqnarray}
C_1 &=& \sum_k k b_k,\nonumber \\
C_2 &=& -\beta {a_0 \over 2} - \sum_k \beta a_k. 
\end{eqnarray}
This completes the solution written down in eq.~(\ref{eq:crux}). Given
the angular structure of the density $G(\theta)$, we can find that of
the potential $F(\theta)$, and vice versa.

\section{Fourier Coefficients of Elliptical Distributions}

Let us consider the Fourier expansion of the elliptical distribution
in the flat rotation curve case ($\beta = 1$)
\begin{equation}
F(\theta) = A(\cos^2\theta + q^{-2} \sin^2\theta)^{-1/2}.
\label{eq:ellipse}
\end{equation}
The Fourier coefficients are
\begin{eqnarray}
a_{k} &=& {1 \over \pi} \int_0^{2\pi} F(\theta) \cos(k\theta)
d\theta \\
b_{k} &=& {1 \over \pi} \int_0^{2\pi} F(\theta) \sin(k\theta)
d\theta.
\end{eqnarray}
However, the only non-vanishing Fourier coefficients are $a_{2n}$
where $n$ is an integer. The lowest-order terms are
\begin{eqnarray}
a_0 & = & {4A\over \pi} K\nonumber \\
a_2 & = & {4A\over \pi (1\!-\!q^2)}[(1\!+\!q^2)K\!-\!2q^2 E] \nonumber \\
a_4 & = & {4A\over 3\pi (1\!-\!q^2)^2}[(3q^2\!+\!1)(q^2\!+\!3)K 
\!-\!8q^2 (1\!+\!q^2)E] \nonumber \\
a_6 & = & {4A\over 15\pi
(1\!-\!q^2)^3}[(1\!+\!q^2)(15\!+\!98q^2\!+\!15q^4)K \\
    &   & \qquad\qquad\qquad - 2q^2(23\!+\!82q^2\!+\!23q^4)E] \nonumber \\
a_8 & =& {4A \over 105\pi (1\!-\!q^2)^4} [
-32q^2(1\!+\!q^2)(11\!+\!74q^2\!+\!11q^4)E \nonumber \\
    &  &  +(105\!+\!1436q^2\!+\!3062q^4\!+\!1436q^6\!+\!105q^8)K],\nonumber
\end{eqnarray}
where $K$ and $E$ denote the following integrals
\begin{eqnarray}
K & =& \int_0^{\pi/2} {d \vartheta \over (1 -
(1-q^{-2})\sin^2\vartheta)^{1/2}} \nonumber\\
E & =& \int_0^{\pi/2} d \vartheta (1 - (1-q^{-2})
\sin^2\vartheta)^{1/2}.
\end{eqnarray}
For elliptical potential models, this gives the Fourier series for the
angular function $F(\theta)$.  Note that the expressions become more
and more cumbersome with increasing order. However, the Fourier series
converges (numerically) extremely rapidly.

For elliptical density models, exactly the same coefficients hold good
but for the Fourier expansion for $G(\theta)$, the shape function in
the density. Of course, the Fourier coefficients of $G(\theta)$ are
directly related to the Fourier coefficients of the potential contour
$F(\theta)$ via eqs. (\ref{eq:Fourier}) and (\ref{eq:Fourieragain}).

\section{Calculation of the Isophotal Distortions}

Most giant elliptical galaxies have isophotes that are very well
approximated by ellipses. The same is of course not true for disk or
barred or interacting galaxies. Data from numerical simulations of
halo formation are sparse, but suggest that dark haloes may also be
well-approximated by ellipses (e.g., Heyl et al. 1994).  So, it is
useful to be able to estimate the size of the distortions of our fits
from pure ellipses. This is done by computing Bender et al.'s (1989)
coefficients, defined as
\begin{eqnarray}
a_k^B &=& {1\over \pi} \int_0^{2\pi} \left(\ri(\theta) - \re (\theta)
\right) \cos (k \theta), \\
b_k^B &=& {1\over \pi} \int_0^{2\pi} \left(\ri(\theta) - \re (\theta)
\right) \sin (k \theta),
\end{eqnarray}
where $\ri$ is the polar equation of the isophote and $\re$ is the
best fitting ellipse (see also Binney \& Merrifield 1998). In
particular, a positive (or negative) $a_4^B$ means that the isophotes
are disk-like (or boxy).

From eqn~(\ref{eq:pois}), we deduce the equation of the isophote as
\begin{equation}
\ri = { a_0 \over 2 }\!+\!\sum_{k = 2}^{\infty} [a_k 
(1\!-\!k^2) \cos(k\theta)\!+\!b_k (1\!-\!k^2) \sin(k
\theta)].
\end{equation}
Here, we have specialised to the case of a flat rotation curve by
setting $\beta =1$.  As shown in Section 3.3, we can transform to the
coordinate system aligned with the major and minor axis and so we can
set $b_2 =0$.

Now we must determine the best fitting ellipse, which is the one which
reproduces the same Fourier coefficients $a_0$ and $a_2$ as our
isophote.  In other words, we solve for the amplitude $A$ and axis
ratio $q$ in eqn~(\ref{eq:ellipse}),
\begin{eqnarray}
a_0 & = & {4A\over \pi} K,\nonumber \\
-3a_2 & = & {4A\over \pi (1\!-\!q^2)}[(1\!+\!q^2)K\!-\!2q^2 E].
\end{eqnarray}
These equations do not have analytic solutions, but the roots can be
easily obtained numerically. Having solved for $A$ and $q$, our
isophote $\ri$ and the best fitting ellipse $\re$ have Fourier series
which differ only for coefficients $a_k$ and $b_k$ with $k\ge 3$.

It is now straightforward to derive formulae for Bender et al.'s
coefficients. Tabulated values are available in the literature only
for the $a_3^B$ and $a_4^B$, so we confine explicit formulae to these
cases:
\begin{eqnarray}
a_3^B &=& -8a_3, \\ 
a_4^B &=& -15a_4\!-\!{4A\over 3\pi
(1\!-\!q^2)^2}[(3q^2\!+\!1)(q^2\!+\!3)K \!-\!8q^2 (1\!+\!q^2)E]
\nonumber 
\end{eqnarray}
Numerical values for the lens galaxies in Q\,2237+030 PG\,1115+080 are
given in the main text. In both cases, the amplitudes of $a_3^B$ and
$a_4^B$ are within the expected ranges, confirming that our solutions
are realistic.

\section{Inclusion of External Shear}

In this appendix, we give the equations incorporating the effects of
external shear. The lens equation becomes
\begin{equation}
\xi = x\!+\!\gamma_1 x\!+\!\gamma_2 y\!-\!{\partial \phi \over \partial x}, 
\qquad
\eta = y\!+\!\gamma_2 x\!-\!\gamma_1 y\!-\!{\partial \phi \over \partial y}.
\end{equation}
The vector of observables becomes
\begin{eqnarray}
{\bd} &=& \Bigl( (1\!+\!\gamma_1) r_\ell \cos\theta_\ell + \gamma_2 r_\ell \sin
\theta_\ell,\dots, \nonumber \\
& & (1\!-\!\gamma_1) r_\ell \sin\theta_\ell + \gamma_2
r_\ell\cos\theta_\ell, \dots, \nonumber\\  
& & (f_{1\ell}\!-\!1)r_1 r_\ell
(1\!-\!\gamma_1^2\!-\!\gamma_2^2), \dots \Bigr)^T,
\label{eq:observablesnew}
\end{eqnarray}
where $\ell$ runs from $1$ to $n$ for a lens system with $n$ images.
The matrix $C$ retains the same form as given in eq~(\ref{eq:defC}),
but the coefficients $\gamma_k\ell$ and $\delta_{k\ell}$ become
\begin{eqnarray}
\gamma_k (\theta_\ell) &=& (1\!-\!k^2)
[f_{1\ell} r_\ell \cos(k \theta_1)W(\theta_1)\!-\!r_1 \cos(k \theta_\ell)W(\theta_\ell)]
\nonumber \\
\delta_k (\theta_\ell) &=&  (1\!-\!k^2)
[f_{1\ell} r_\ell \sin(k \theta_1)W(\theta_1)\!-\!r_1 \sin(k
\theta_\ell)W(\theta_\ell)] \nonumber,
\end{eqnarray}
with
\begin{equation}
W(\theta) = 1\!+\!\gamma_1 \cos2\theta_\ell\!+\!\gamma_2 \sin 2\theta_\ell.
\end{equation}
The vector of unknowns is unchanged from eq~(\ref{eq:vectorx}).
For a given external shear, the problem remains linear and can be
solved by SVD. 

\end{appendix}

\label{lastpage}

\end{document}